\documentclass[%
 reprint,
 amsmath,amssymb,
 aps,pre
]{revtex4-2}

\usepackage[dvipsnames]{xcolor}
\usepackage{graphicx}
\usepackage{dcolumn}
\usepackage{bm}
\usepackage{enumitem}
\usepackage{mathtools,physics}
\usepackage{tikz-cd,tikz}
\usepackage{mdframed}

\newcommand{\Reals}{\mathbb{R}}

\newcommand{\weakp}[1]{\langle\!\langle #1 \rangle\!\rangle}
\newcommand{\innerp}[1]{\langle #1 \rangle}
\newcommand{\inv}{^{-1}}
\newcommand{\T}{^{\top}}
\newcommand{\half}{\frac{1}{2}}

\newtheorem{definition}{Definition}

\DeclareRobustCommand{\firstedit}{}

\begin{document}

\preprint{APS/123-QED}

\title{Contraction and Synchronization in Reservoir Systems}
\thanks{Distribution Statement A: Approved for Public Release; \\Distribution is Unlimited. PA\# AFRL-2024-2340}

\author{Adrian S. Wong}
\email{adrian.wong.ctr@us.af.mil}

\affiliation{%
	Jacobs Technology Inc.\\
	and\\
	In-space Propulsion Branch\\
	Air Force Research Laboratory \\
	Edwards AFB, CA 93524
}%

\author{Robert S. Martin}
\affiliation{%
	Army Research Office \\
	DEVCOM Army Research Laboratory\\
	Durham, NC 27703 
}%

\date{\today}

\author{Daniel Q. Eckhardt}
\affiliation{%
	In-space Propulsion Branch\\
	Air Force Research Laboratory \\
	Edwards AFB, CA 93524 
}%

\date{\today}

\begin{abstract}
This paper explores the conditions under which global contraction manifests in the leaky continuous time reservoirs, thus guaranteeing generalized synchronization. Results on continuous time reservoirs make use of the logarithmic norm of the connectivity matrix. Further analysis yields some simple guidelines on how to better construct the connectivity matrix in these systems. Additionally, we outline how the universal approximation property of discrete time reservoirs is readily satisfied by virtue of the activation function being contracting, and how continuous time reservoirs may inherit a limited form of universal approximation due to their overlap with neural ordinary differential equations. The ability of the reservoir computing framework to universally approximate topological conjugates, along with their fast training, make them a compelling data-driven, black-box surrogate of dynamical systems, and a potential candidate for a component of digital twins. 

\begin{description}
\item[Keywords]
reservoir computing, contraction theory, recurrent neural networks, synchronization, \\topological conjugacy, digital twins
\end{description}
\end{abstract}

\maketitle


\section{Introduction}

Echo state networks (ESN) and liquid state machines (LSM) were proposed as promising architectures for neural networks more than two decades ago. The ESN approach comes from the perspective of artificial neural networks using a discrete time formulation and sigmoid activation function \cite{Jaeger2001}. The LSM approach, on the other hand, comes (more directly) from the perspective of biological neural networks, where the network vector field follows the integrate-and-fire neuron model in continuous time \cite{Maass2002}. Soon after, the term \textit{reservoir} was first invoked in \citet{Steil2004}, and the term \textit{Reservoir Computing} (RC) was coined in \citet{Verstraeten2005} while attempting to implement these architectures in field-programmable gated arrays.

Before continuing further, we define some terms to avoid potential confusion. We use the term \textit{reservoir computing} to mean the series of steps used to set up and train a recurrent neural network, which is treated as a driven dynamical system. The term \textit{reservoir} refers to the driven state-space model; it has fixed or prescribed structure and has not yet been trained. The \textit{trained} or \textit{autonomous} reservoir refers to the autonomous state-space model, which incorporates the trained structure from a readout \textit{layer}. As such, RC is a framework that utilizes the dynamics of an arbitrarily prescribed reservoir, where the training is performed only on the readout layer. 

\firstedit{For the RC to have expressiveness in the space of nonlinear dynamics, either the reservoir dynamics or the readout must contain nonlinearity. Linear reservoir dynamics must be accompanied by a nonlinear readout, but nonlinear reservoir dynamics may be accompanied by either a linear or nonlinear readout. Most commonly, nonlinear reservoir dynamics are paired with a linear readout which is the case that this paper will focus on. RC with a nonlinear-nonlinear pairing has shown some success, but so far has not seen widespread usage \cite{Pathak2018}. }

The RC framework results in recurrent networks with some extremely promising aspects. It is able to predict autonomous dynamical systems with higher accuracy and smaller network sizes than deep feedforward networks and \textit{long-short term memory} (LSTM) recurrent networks \cite{Chattopadhyay2020}. The training process is noniterative thus quick -- if the dynamics are nonlinear, the training requires only a linear fit, compounding on the aforementioned benefits of RC. This is therefore a promising framework, with compelling reasons to understand its theoretical underpinnings and the source of its many strengths. This has yielded considerable research efforts since its conception, particularly in the role of contraction in the \textit{listening phase} of RC \cite{Yildiz2012, Buehner2006, Zhang2011, Manjunath2013,Grigoryeva2018}. 

More recently, \citet{Pathak2017} studied RC in the context of replicating the Lyapunov exponents of chaotic systems, followed by \citet{Lu2018} studying the reconstruction of attractors with reservoirs. Even though \citet{Jaeger2001} and \citet{Steil2004} demonstrated successful prediction on the chaotic Mackey-Glass system, \citet{Pathak2018} showed the astounding ability of autonomous reservoirs as predictive models, which made waves in the chaotic dynamics community. \citet{Lu2018} then discussed the link between the echo state property (ESP) of the listening phase and the phenomenon of generalized synchronization (GS) -- a well-studied phenomena in chaotic systems \cite{kocarev1996generalized,Rulkov1995,Abarbanel1996}. Together, these papers suggest that there was substantial overlap between chaotic dynamics and reservoirs. The relationship between ESP and GS was further studied and formulated in more rigorous terms by \citet{hart2020embedding}. 

The recognition that GS and the ESP were closely related phenomena was a particularly fruitful realization \cite{grigoryeva2021chaos, Grigoryeva2023, hart2023, Lymburn2019, Verzelli2021} and one of the core motivations of this research effort. In this paper, we attempt to connect yet another concept -- contraction theory \cite{Lohmiller1998, aminzare2014synchronization, aminzare2014contraction, bullo2023, sontag2010contractive} -- with GS and the ESP. The ESP in continuous time reservoirs is relatively understudied compared to their discrete time counterparts. We give a preamble with discrete time reservoirs and ESP before pivoting to continuous time reservoirs. As far as the authors are aware of, \citet{hart2023} provides the only results on the sufficient conditions of continuous time ESP. Our approach differs from that of \citet{hart2023} but yields similar results that can be quickly computed.

The contraction that we invoke is better described as \textit{globally uniform state-contraction}. Even though \textit{locally uniform state-contraction} is sufficient for the ESP, the claims that we set out to make are nontrivial when considering all possible local cases, so we resort to the narrower global case. From here on, \textbf{the prefix \textit{globally uniform state-} is always implied when mentioning \textit{contraction}}, unless otherwise specified. It should also be understood that all inputs $u$ to the reservoir are measurements of an autonomous dynamical system with invertible velocity field and bounded trajectories\firstedit{, and that the description \textit{smooth} will always mean \textit{continuously differentiable}.}

It is important to keep in mind that ESP and GS are only necessary conditions for the successful implementation of the RC framework; \textbf{it guarantees very little about the resulting autonomous reservoir and the quality of its predictions}. Even the universal approximation property (UAP) is not necessary, but a welcomed property -- when possible -- for a successfully trained reservoir to be a universal approximation of autonomous dynamical systems. 

The organization of this paper is as follows. Section \ref{sec:disc-time-systems} goes over the basic structure of discrete time reservoirs. It gives the current understanding of the role of contraction and displays how certain contracting properties can be leveraged. Section \ref{sec:cont-time-systems} handles contraction in continuous time reservoirs, which bears significant resemblance to the approach in Section \ref{sec:disc-time-systems}. Our results provide a quickly computable sufficient condition that can be easily adjusted to guarantee contraction. Section \ref{sec:discussions} discusses the UAP in the context of contracting activation functions, and topological conjugacy of the autonomous reservoir. Lastly, to avoid lengthy tangents, the appendices cover some specific details and claims made in the main sections.

\section{Discrete Time Systems}\label{sec:disc-time-systems}

Suppose we are given some bounded and time varying input signal $u = \{u_t : t=0,1,\cdots,T\}$ that are measurements of some dynamical system, where each $u_t\in \Reals^M$. This input signal $u_t$ is used to drive two identical driven dynamical systems with $N$ nodes, differing only in their initial conditions $x_{0} \ne y_{0}$, imposing for now that $N\gg M$. Two trajectories $x = \{x_t\}$ and $y = \{y_t\}$, both of length $T+1$, are generated as a result of the following vector equations, with states $x_t,y_t \in \Reals^N$. 
\begin{equation}
	\begin{aligned}
		x_{t+1} = \sigma(Ax_t+Bu_t), \\
		y_{t+1} = \sigma(Ay_t+Bu_t). \label{eq:dynamics1}
	\end{aligned}
\end{equation}

These driven dynamical systems described in \eqref{eq:dynamics1} are called \textit{reservoirs} with $N$ nodes in the context of this paper. The connectivity matrix $A\in\Reals^{N\times N}$ describes the connections between the reservoir nodes. Matrix $B\in\Reals^{N\times M}$ is the input gain of the reservoir, designating how much each node is coupled to the inputs. The activation function, typically the hyperbolic tangent function, is the same for every node and is applied component-wise to its vector argument, i.e. $\sigma : \Reals^N\to\Reals^N$ such that $\sigma([v_1,v_2]\T)=[\sigma(v_1),\sigma(v_2)]\T$. 

It is not expected that these two systems will reach a static equilibrium, since they are both driven by a time varying input $u$. Instead, consider defining a \textit{virtual displacement system} (VDS) as $x_{t}-y_{t}$ to understand the long-term behavior of the individual systems \cite{Lohmiller1998}. Specifically, it is sufficient that if the VDS shrinks according to some norm for every $t \ge 0$, the two parent systems in \eqref{eq:dynamics1} will asymptotically approach the same solution and be identically synchronized, which implies that the the reservoir is synchronized (in the sense of GS) to the dynamical systems generating the inputs \cite{kocarev1996generalized, Abarbanel1996}:
\begin{equation}
	\norm{ x_{t+1}-y_{t+1} } \le k\norm{ x_{t}-y_{t} }. \label{eq:exp-stable}
\end{equation}

For some vector $v$, let $\norm{v}$ denote any vector norm unless otherwise specified. The property described by \eqref{eq:exp-stable} yields \textit{exponential stability} with constant $0 \le k < 1$, which also guarantees \textit{asymptotic stability}. To demonstrate exponential stability, we make use of the contraction mapping property of the sigmoid family of functions for any two vectors $a\ne b$ in $\Reals^N$:
\begin{equation}
	\norm{\sigma(a)-\sigma(b)} \le \norm{ a-b }. \label{eq:contracting}
\end{equation}

In one dimension, contractions guarantee that two distinct points on its curve cannot generate a slope greater than unity and is equivalent to limiting the supremum magnitude of the derivative of $\sigma$ to unity. In higher dimensions, a contracting $\sigma$ can only shorten the distance between two vectors, even if a shift is applied. For some matrix $M$, let $\norm{M}$ denote any induced matrix norm. The matrix and vector norms throughout this paper will be implied through consistent usage of upper- and lower-case letters of the respective arguments. Using \eqref{eq:contracting}, we show that the VDS is \textit{globally exponentially stable} when driven by any bounded $u$, so long as the induced matrix norm $\norm{A} < 1$. See Appendix \ref{sec:contraction-theory} for additional details.
\begin{equation}
	\begin{aligned}
		\norm{\sigma(Ax_t+Bu_t)-\sigma(Ay_t+Bu_t)} 	
		&\le \norm{ A(x_t-y_t) } \\
		&\le \norm{A}\!\cdot\!\norm{x_t-y_t} \\
		&< \norm{x_t-y_t}. \label{eq:discrete-contraction}
	\end{aligned}
\end{equation}

Notably, the 2-norm is often used, which induces a matrix norm equal to the largest singular value of that matrix. This is in contrast to the common practice of setting the spectral radius of $A$ to be $\rho(A)<1$, as recommended by \citet{Zhang2011}. We also remark that $\rho(A)<1$ does not guarantee that $\norm{A} < 1$ since $\rho(A)\le \norm{A}$. This result is more general in that it applies to all norms, so if just one of the norms results in contraction, then contraction in all other norms will eventually follow. 

Defining in this context $\norm{A}_\ell$ as some arbitrary norm, a broader sufficient condition for stability is $\min_\ell \norm{A}_\ell < 1$. For ease of calculation, $\ell$ is usually chosen to correspond to the $1$-, $2$-, or $\infty$-norm. We have shown that when $\norm{A} < 1$, the systems in \eqref{eq:dynamics1} will uniformly and exponentially approach one another at a rate no slower than $\norm{A}$ regardless of initial conditions. If $A$ can be prescribed, then a reservoir can be constructed such that the VDS is always globally exponentially stable:
\begin{equation}
	\norm{ x_{t}-y_{t} } \le \norm{A}^t \norm{ x_{0}-y_{0} } \Longrightarrow \lim_{t\to\infty}{ x_{t} - y_{t} } = 0.
\end{equation}

The rate at which the distance shrinks depend on the choice of norm. Also notice that the right-hand side of \eqref{eq:discrete-contraction} is independent of $u_t$ and $B$. All this implies that a particular reservoir, defined by the triplet $(A,B,\sigma)$ with $\norm{A} < 1$, will map every point of the input trajectory $u=\{u_t\}$ to a unique point in the reservoir trajectory $x=\{x_t\}$ in the infinite time limit \cite{grigoryeva2021chaos, kocarev1996generalized}. Necessarily, the ESP holds, and a unique GS occurs between the drive and response system:
\begin{equation}
	\lim_{t\to\infty}\frac{1}{t}\log\frac{\norm{ x_{t}-y_{t} }}{\norm{ x_{0}-y_{0} } } = \Lambda \le \log \norm{A} < 0. \label{eq:CLE1}
\end{equation}

This condition on $A$ is closely related to the largest \textit{conditional Lyapunov exponent} (CLE) of the system $\Lambda$, conditioned on the driving signal $u$. In fact, $\log \norm{A}$ is the negative-definite upper bound of $\Lambda$, which is itself the upper bound on all the other CLEs. This implies GS, so the reservoir implicitly has a globally attractive trajectory $x=\{x_t\}$ unique to any given $u = \{u_t\}$ as $t\to\infty$. In the infinite time limit, there exists a unique \firstedit{and smooth} function $\Phi : \Reals^M\to\Reals^N $, corresponding to the triplet $(A,B,\sigma)$ with constraint $\norm{A}\le 1$, that maps the driving signal to the response, i.e. $x_t=\Phi(u_t)$ as $t\to\infty$. \firstedit{The smoothness of $\Phi$ can be shown using Theorem 2.2.2 of \citet{hart2020embedding}.} For any desired error $\varepsilon > 0$ and any chosen norm, there exist some function $\Phi$ and some time $T>0$ such that $\norm{x_{T}-\Phi(u_T)} < \varepsilon$.

\section{Continuous Time Systems}\label{sec:cont-time-systems}

\subsection{Similarity to Discrete Time}\label{sec:cont-RC}
In continuous time, the following situation is considered instead. Suppose that we are given some bounded and time varying input signal $u(t)\in\Reals^M$ in the time window $t\in[0,T]$ that is driving two identical reservoirs in the vector field according to the equations below:
\begin{equation}
	\begin{aligned}
		\dot x = -Cx+\sigma(Ax+Bu), \\
		\dot y = -Cy+\sigma(Ay+Bu). \label{eq:dynamics2}
	\end{aligned}
\end{equation}

As before, these two reservoirs have states $x(t)\in\Reals^N$ and $y(t)\in\Reals^N$ that stem from different initial conditions $x(0)\ne y(0)$. The matrices $A\in\Reals^{N\times N}$ and $B\in\Reals^{N\times M}$ serve the same purposes as the discrete time case, but an additional \textit{leak} matrix $C\in\Reals^{N\times N}$ is necessary for stability, as will be demonstrated. The term \textit{leak} is in reference to the leak current in biological neurons. For mathematical generality (as compared to biological realism), we impose that $C$ be \textit{symmetric positive definite} (SPD). The activation function $\sigma$ is from the usual sigmoid family, applied component-wise as before. Define the VDS by $z(t)\equiv x(t)-y(t)$, with dynamics given by the following equation:
\begin{equation}
	\dot z = -C(x-y)+\sigma(Ax+Bu)-\sigma(Ay+Bu). 
\end{equation}

A \textit{Lyapunov function} $V(z)=\half\innerp{ z,z} = \half \norm{z}_2^2$ can be defined for $z$, where the evolution of this function is $\dot V(\dot z,z)= \innerp{ \dot z,z}$. The function $V(z)$ and its evolution $\dot V(\dot z,z)$ can be acquired explicitly through inner products with $z$. 
\begin{equation}
	\innerp{ \dot z, z} = \underbrace{-\innerp{ Cz,z}}_{\le -s_{N}(C) z^2} +\underbrace{\innerp{\sigma(Ax+Bu)-\sigma(Ay+Bu), x-y}}_{\le s_{1}(A) z^2}. \label{eq:continuous-contraction}
\end{equation}

The first RHS term is bounded from above by $-s_{N}(C) z^2$, where $s_{N}(C)$ denotes the $N^{\text{th}}$ (smallest) singular value of $C$. See Appendix \ref{sec:spd-proof} for the proof of this inequality. Using the contraction of $\sigma$, the second RHS term is bounded from above by $s_{1}(A) z^2$ instead, where $s_{1}(A) = \norm{A}_2 $ is the first (largest) singular value of $A$. See \eqref{eq:contraction4} for derivation of this inequality. Now define $k= (s_{N}(C) - s_1(A))/2$. As long as the matrices $A$ and $C$ are such that $k > 0$, we have a negative definite $\dot V(\dot z,z)$ away from the unique stable equilibrium at zero. The VDS approaches the unique equilibrium at the minimum rate of $k$, where $\dot V(\dot z,z)=0$ if and only if $z=0$. Similarly, $V(z)=0$ if and only if $z=0$ and is otherwise positive, by definition. It should be noted that \textit{equilibrium} here does not mean that $x(t)$ or $y(t)$ approach a static value, merely that their difference $z(t)$ does. 
\begin{equation}
	\dot V(\dot z,z) \le -2k V(z) \Longrightarrow \norm{z(t)}_2 \le \norm{z(0)}_2 e^{-kt}.\label{eq:Lyap-func1}
\end{equation}

Given the evolution of this Lyapunov function, $k > 0$ results in the VDS being \textit{globally exponentially stable}, making the parent systems \eqref{eq:dynamics2} identically synchronized to one another, and the reservoir generally synchronized to the input. Since the evolution $\dot V(\dot z,z)$ is ultimately a function of $z$ and $u$, $V(z)$, it is a \textit{control-Lyapunov function} rather than a simple Lyapunov function. This is an important distinction as it suggests that $u$ plays a pivotal role in determining the attracting manifold, but plays no role determining whether the manifold \textit{is} attracting. It is perhaps also a partial explanation as to why reservoirs robustly adapt to many input signals. Similar to the discrete time case, $-k$ is the upper bound of the largest CLE of the system:
\begin{equation}
	\lim_{t\to\infty}\frac{1}{t}\log\frac{\norm{ z(t) }_2}{\norm{ z(0) }_2 } = \Lambda \le -k < 0 .\label{eq:CLE2}
\end{equation}

In the asymptotic limit, the attracting trajectory of $x(t)$ is unique for a given input $u(t)$, up to the quadruplet $(A,B,C,\sigma)$ and the aforementioned constraints on $A$ and $C$. It is common in practice to use the identity matrix $I$ instead of the generic SPD matrix $C$. Of course, the identity matrix is itself SPD, but this fixes $s_{N}(I) = 1$ and restricts $s_{1}(A) = \norm{A}_2 < 1$, displaying contraction with same triplet $(A,B,\sigma)$ as the 2-norm discrete time case, but with leaky integrate-and-fire dynamics. In the infinite time limit, there exists a \firstedit{unique and smooth} function $\Phi : \Reals^M\to\Reals^N $ that maps $u(t)$ to $x(t)$, i.e. $x(t)=\Phi(u(t))$ as $t\to\infty$ \cite{ hart2020embedding, sauer1991embedology}. \firstedit{Though \citet{hart2020embedding} handles the discrete time case, it makes use of contraction to show smoothness and uniqueness. We have shown that contraction occurs in the continuous time case, so $\Phi$ will similarly be smooth and unique.} For any tolerance $\varepsilon > 0$, there exist some function $\Phi$ and time $T>0$ such that $\norm{x(T)-\Phi(u(T))}_2 < \varepsilon$. 


\subsection{Generalizing with Weak Pairings}\label{sec:weak-pair}

Though we have shown that $\norm{A}_2 < 1$ results in exponential stability using an inner product, contraction of these continuous time systems can be better demonstrated in continuous time systems through other concepts \cite{Qiao2001}. Consider a generalization of an inner product called a \textit{weak pairing}, closely related to the semi-inner product (in the sense of Lumer \cite{Lumer1961}). We remark that this semi-inner product forgoes the symmetry of the arguments and is unrelated to the seminorm. 

Define the weak pairing $\weakp{ \cdot\, , \cdot } : \Reals^N\times\Reals^N \to \Reals$ as a nonnegative map, with properties defined in Appendix \ref{sec:weak-pairing-prop}. For any vector norm $\norm{\cdot}$, there exists a compatible (but not necessarily unique) weak pairing $\weakp{ \cdot,\cdot }$ such that $\weakp{ v,v} = \norm{v}^2$ for all $v\in\Reals^N$. If the norm is induced by an inner product, then the weak pairing and an inner product coincide. Importantly, these weak pairings are closely related to \textit{logarithmic norms} (henceforth lognorms) $\mu(A)$ and satisfy \textit{Lumer's equality}:
\begin{equation}
		\mu(A) = \max_{\norm{v}=1} \weakp{ Av, v}. \label{eq:log-norm2}
\end{equation}

We will not discuss weak pairings nor lognorms in great depth, only the most relevant properties. The basic properties of lognorms and weak pairings are defined in Appendixes \ref{sec:log-norm-prop} and \ref{sec:weak-pairing-prop} respectively. More detailed treatment can be found in the references therein. 

Broadly speaking, the lognorms $\mu(A)$ will be used to provide bounds for the weak pairings as well as feasible methods of calculation. These weak pairings allow for the definition of a family of candidate Lyapunov functions on the VDS of \eqref{eq:dynamics2}, indexed by $\ell\in\Reals^{\ge1}$, as $\{V_\ell(z):V_\ell(z) = \half\weakp{ z, z}_\ell \ge 0 \}$ with the respective evolution $\dot V_\ell(\dot z,z)= \weakp{ \dot z, z}_\ell$. 
If any of these weak pairings yields $\dot V_\ell(\dot z,z)\le 0$ for all $z$ and $u$, then that $V_\ell(z)$ is a Lyapunov function and the VDS is stable and GS occurs. 

Additionally, if $\dot V_\ell(0) = 0$ and $\frac{\dot V_\ell(z)}{V_\ell(z)} < 0$ for all nonzero $z$, then the system is globally exponentially stable with a unique equilibrium at $z=0$. The stability condition for the proposed family of Lyapunov functions has the following form, which guarantees global exponential stability:
\begin{equation}
	\begin{aligned}
		&\frac{\dot V(\dot z,z)}{V(z)} = \frac{\weakp{-Cz,z}}{\norm{z}^2}\\ 
		&\qquad+\frac{\weakp{\sigma(Ax+Bu)-\sigma(Ay+Bu), x-y}}{\norm{ x-y}^2} < 0.\label{eq:Lyap-func2}
	\end{aligned}
\end{equation}

The first weak pairing term above can be bounded by $\mu(-C)$ by definition. Using the shorthand $\sigma(Ax+Bu) = \hat\sigma(x)$ for the activation function and substitution $x\to y+hv$ for conciseness, the second weak pairing term of \eqref{eq:Lyap-func2} can be bounded by the maximal lognorm over all possible Jacobians $D\hat\sigma(x)$ of the activation function:
\begin{equation}
	\begin{aligned}
		\max_{x\ne y} &\frac{\weakp{ \hat\sigma(x)-\hat\sigma(y), x-y }}{\norm{x-y}^2} \\ 
		&= \max_{y;\norm{v}=1} \lim_{h\to 0^+} \frac{\weakp{ \hat\sigma(y+hv)-\hat\sigma(y), hv }}{h^2 \norm{v}^2}\\
		&= \max_{y;\norm{v}=1} \biggl\langle\hspace{-1em}\biggl\langle \; \lim_{h\to 0^+} \frac{\hat\sigma(y+hv)-\hat\sigma(y)}{h}, v \,\biggr\rangle\hspace{-1em}\biggr\rangle\\ 
		&= \max_{y;\norm{v}=1} \weakp{ D\hat\sigma(y)v, v }\\
		&= \max_{x} \mu(D\hat\sigma(x)). \label{eq:contraction5}
	\end{aligned}
\end{equation}

The sufficient condition for global exponential stability of the VDS is $\max_{x} \mu(D\hat\sigma(x))<-\mu(-C)$. The calculation of this quantity is difficult for arbitrary norms, except for the special cases of the 1-lognorm and the $\infty$-lognorm (that is, $\ell=1$ and $\ell=\infty$) with saturating activation function $\sigma$ (such as tanh). These case admit a closed form solution in terms of the connectivity matrix $A$ \cite{Qiao2001}. The maximum of these lognorms over all possible Jacobians $D\hat\sigma(x)$ yields the following analytic form, with the usage of the entry-wise or Hadamard product $(I\circ A)_{ij} = (I)_{ij}(A)_{ij}$. The proofs and details of this are in Appendix \ref{sec:max-jac}:
\begin{equation}
	\begin{aligned}
		\max_{x} \mu_1(D\hat\sigma(x)) = \max \{ \mu_1(A),\mu_1(A-I\circ A)\},\\
		\max_{x} \mu_\infty(D\hat\sigma(x)) = \max \{ \mu_\infty(A),0\}. \label{eq:cont-esp} 
	\end{aligned}
\end{equation}

It suffices that imposing either $\max \{ \mu_1(A),\mu_1(A-I\circ A)\}<-\mu(-C)$ or $\max\{\mu_\infty(A),0\} <-\mu(-C)$ is enough to guarantee the ESP and GS. These conditions may be readily calculated according to the definitions of $\mu_1(A)$ and $\mu_\infty(A)$ in \eqref{eq:log-norms1} and \eqref{eq:log-norms2} respectively. These lognorms provide a tighter bound than their matrix norm counterparts $\norm{A}_1$ and $\norm{A}_\infty$ respectively, as shown at the end of Appendix \ref{sec:max-jac}. If $C$ is chosen to be the identity, these conditions simplify and are given by the following:
\begin{equation}
	\begin{aligned}
		\max \{ \mu_1(A),\mu_1(A-I\circ A)\} < 1,\\
		\mu_\infty(A) < 1. \label{eq:cont-esp2}
	\end{aligned}
\end{equation}

Only the 1-lognorm and $\infty$-lognorm cases are checked in the approach detailed here. When neither of these conditions are satisfied, it is still possible that contraction occurs through some other lognorm, but it is not yet known how to acquire those sufficient conditions. From this analysis, it is clear that the leak term $C$ must be present for contraction to be guaranteed with logarithmic norms. This is due to both $\max \{ \mu_1(A),\mu_1(A-I\circ A)\}$ and $\max\{\mu_\infty(A),0\}$ always being nonnegative. The stability provided by the leak term cannot be overlooked if ESP and GS are to be guaranteed with this approach. Nevertheless, these are only sufficient conditions and it is still possible for reservoir systems to show ESP and GS without satisfying these conditions. 

\subsection{Interpretation}

Though these results appear somewhat esoteric, there are potential interpretations of these results. The previous section showed that if either of the two conditions in \eqref{eq:cont-esp2} are satisfied, the ESP and GS of the reservoir is guaranteed. These conditions have implications on the structure of the connectivity matrix $A$. 

The first condition, corresponding to the 1-lognorm in \eqref{eq:cont-esp2}, can be further broken down into two parts -- $\mu_1(A)< 1$ and $\mu_1(A-I\circ A) < 1$. Both subconditions must be true for the condition to hold. The $1$-lognorm of an arbitrary matrix $M$ can be calculated via the following:
\begin{equation}
	\mu_1(M) = \max_{j} \left( m_{jj} + \sum_{i\ne j}\abs{m_{ij}} \right). \label{eq:log-norms1}
\end{equation}

For the connectivity matrix $A$, $\mu_1(A)< 1$ can be easily calculated according to \eqref{eq:log-norms1}, where the column-sum is over the the magnitude of the off-diagonal terms. Effectively, this imposes that for every column (indexed by $j$), the sum the off-diagonal $\abs{m_{ij}}$ must be less than $1-a_{ij}$:
\begin{equation}
	 \sum_{i\ne j}\abs{a_{ij}} < 1 - a_{jj}. \label{eq:subcond1}
\end{equation}

This is consistent with the common practice of choosing $A$ to be sparse, since the $\sum_{i\ne j}\abs{a_{ij}}$ term grows quickly with a denser $A$. This condition suggests that a more negative diagonal element is desirable, which ties in the concept of dissipation and its role in synchronization\cite{Pereira2014, aminzare2014synchronization, Hale1997}. In addition, this condition fails to hold when $a_{jj}\ge 1$, due to the sum of absolute values always being nonnegative. 

The second subcondition $\mu_1(A-I\circ A) < 1$ appears more complicated, but is conceptually simple as well. Let the Hadamard product be represented by the operator $\circ$, so $I\circ A$ constructs a diagonal matrix with the diagonal elements of $A$. The resulting matrix $A-I\circ A$ is similar to $A$, but with its diagonal elements set to zero. This subcondition is a special case of \eqref{eq:subcond1}, given by imposing the following constraint for every column:
\begin{equation}
	\sum_{i\ne j}\abs{a_{ij}} < 1. \label{eq:subcond2}
\end{equation}

Though \eqref{eq:subcond1} and \eqref{eq:subcond2} appear to overlap, these two subconditions disagree when the diagonal elements $a_{jj}$ are positive. Otherwise, if $a_{jj}$ are all nonpositive, then \eqref{eq:subcond2} implies \eqref{eq:subcond1}. 

The second condition for stability is $\max \{ \mu_\infty(A),0\} < 1 $. Zero appears in this equation only to indicate that the leak term with matrix $C=I$, discussed in Section \ref{sec:weak-pair}, must be present to guarantee ESP and GS. Adhering to this constraint, the condition is effectively $\mu_\infty(A) < 1$. This corresponds to the $\infty$-lognorm, which for an arbitrary matrix $M$ can be calculated by the following:
\begin{equation}
	\mu_\infty(M) = \max_{i} \left( m_{ii} + \sum_{j\ne i}\abs{m_{ij}} \right).\label{eq:log-norms2}
\end{equation}

The above condition is the row-wise version of \eqref{eq:log-norms1}, giving the corresponding restriction \eqref{eq:subcond1} when applied to the connectivity matrix $A$:
\begin{equation}
	\sum_{j\ne i}\abs{a_{ij}} < 1 - a_{ii}. \label{eq:subcond3}
\end{equation}

All this suggests that more-negative diagonal elements, and smaller (in magnitude) off-diagonal elements, contribute to ESP and GS. Given these relations, one might even consider drawing the diagonal and off-diagonal elements from different distributions. 

We stress that these are the sufficient conditions for contractions, which are themselves the sufficient conditions for GS as outlined in \eqref{eq:CLE1} and \eqref{eq:CLE2}. It could often be the case, depending on how $A$ was generated, that the GS induced by these constraints will be too restrictive so that the predictions are unreliable, as discussed in \citet{Buehner2006} for the discrete time case. 

The necessary conditions for GS can be determined by calculating the CLEs, or by using the auxiliary systems approach \cite{Abarbanel1996}. However, unlike the analysis performed here, neither of these methods will yield any insight on how the connectivity matrix $A$ be constructed.

\section{Discussions}\label{sec:discussions}

\subsection{UAP in Discrete Time}

Taking the discrete time reservoir outlined in \eqref{eq:dynamics1}, we have established that $A : \norm{A} < 1$ and $B$ can be arbitrarily chosen such that the ESP holds. The generation of trajectories is often called the \textit{listening phase} and implies the existence of a unique time-invariant function $\Phi : \Reals^M\to\Reals^N $ such that $x_t = \Phi(u_t)$ as $t\to\infty$, but without an explicit representation of $\Phi$:
\begin{equation}\begin{aligned}
		x_{t+1} &= \sigma(Ax_t+Bu_t),\\
		z_{t} &= Wx_{t}+b.\label{eq:RC-driven}
\end{aligned}\end{equation}

Suppose now that a linear readout $z_t$ is an attempt to reconstruct the inputs $u_{t}$ up to some error, resulting in an approximation $\hat u_t$. The readout $z_t$ is substituted with the reconstruction $\hat{u}_{t}$ from here on. Effectively, this is an assumption that $\Phi$ is invertible and that we are imposing a linear approximation of $\Phi\inv$ where $u_t=\Phi\inv(x_t) \simeq Wx_{t}+b = \hat u_t$. The system \eqref{eq:RC-driven} can be completed as a reservoir by solving for both the \textit{weights} matrix $W\in\Reals^{M\times N}$ and \textit{bias} vector $b\in\Reals^M$ in the linear least squares sense such that $\hat{u}_{t} \simeq {u}_{t}$. This step is generally referred to as \textit{training} and a regularization term for the least squares fit is commonly used. 

The constraint $\norm{A} < 1$ allows for only one basin of attraction in the reservoir, resulting in the map $\Phi$ induced by the dynamics to also be unique. Aside from some transients, $u$ can be recovered from $x$ through the application of $\Phi\inv$, approximated by $W$. Further suppose that the approximation of $\Phi\inv$ by $W$ holds arbitrarily well, such that $u_T = \Phi\inv(x_T) = Wx_T$ for a finite $T$. As such, the second line of \eqref{eq:RC-driven} can be substituted into the first, resulting in an autonomous reservoir that is used for $t>T$. With the shorthand $\bar A = A+BW$ and $\bar b = Bb$, this newly formulated and autonomous reservoir is usually called the \textit{predicting phase}:
\begin{equation}\begin{aligned}
		x_{t+1} &= \sigma(Ax_{t}+BWx_{t}+{Bb})\\
		&= \sigma(\bar Ax_{t}+\bar b),\\
		\hat u_{t} &= Wx_{t}+b.\label{eq:RC-autonomous}
\end{aligned}\end{equation}

It is surprising that a nonautonomous system like \eqref{eq:RC-driven} can be converted into an autonomous system with a linear readout in the manner described in \eqref{eq:RC-autonomous}. Often overlooked is that the form of \eqref{eq:RC-autonomous} admits the UAP in the feedforward sense of \citet{cybenko1989approximation}, \citet{hornik1989multilayer}, or \citet{leshno1993multilayer}. Approximation theorems for recurrent networks are not necessary. 

Also overlooked is the relationship between contracting functions and the activation functions compatible with the UAP. Nonlinear contracting functions used as activation functions will always be compatible with the UAP. This is most succinctly shown with the results of \citet{leshno1993multilayer} which states that the UAP holds in networks of the form \eqref{eq:RC-autonomous} if and only if the activation function is nonlinear and nonpolynomial. 

The proof of nonlinear contracting functions being nonpolynomials is outlined as follows. If two distinct points on a smooth function cannot generate a slope greater than unity, then this function is globally contracting. Degree one and zero polynomials are linear, so they are automatically excluded from the set of nonlinear contracting functions. Polynomial functions of degree two or higher are nonlinear but will always manifest an unbounded slope eventually, so they cannot be contracting. Thus, the set of nonlinear contracting functions is mutually exclusive with the set of nonlinear polynomial functions. Therefore, all nonlinear contracting functions are nonpolynomials. 

We can then say that if an driven discrete time dynamical system is nonlinear and globally contracting, it will always have the ESP. Associated with this ESP is a unique filter that maps the inputs to the response. When such a system is paired with a linear readout, the combined system will also possess the UAP.

The only invoked property of the activation function $\sigma$ thus far was nonlinear global contraction, which is a special case of Lipschitz continuous. Other commonly invoked properties of sigmoid functions -- monotonicity and saturation -- are not used in this paper, nor in \citet{cybenko1989approximation}. The UAP requires a large but finite number of nodes, hence the condition that $N \gg M$. \firstedit{In the spirit of Whitney (and consequently Takens when appropriate), almost every smooth map $\Phi : \Reals^M\to\Reals^N $ is an embedding when $N > 2 M$. See Theorem 2.2 of \citet{sauer1991embedology} or Conjecture 2.3.4 of \citet{hart2020embedding}. In the large $N$ limit, the approximation of $\Phi\inv(x_t) \simeq Wx_{t}+b$ also holds arbitrarily well for ergodic systems \cite{hart2021echo}.} 

\subsection{UAP in continuous time}

For continuous time reservoirs, let us first consider the Hopfield-like equations outlined in \eqref{eq:dynamics2} with a leak current. We constrain $A$ such that $\norm{A} < 1$ and $C=I$, but $B$ can be arbitrarily chosen such that the ESP holds. This implies the existence of some time-invariant function $\Phi : \Reals^M\to\Reals^N $ where $x(t) = \Phi(u(t))$ as $t\to\infty$ without explicit knowledge about the representation of $\Phi$:
\begin{equation}\begin{aligned}
		\dot x 	&= -x+\sigma(Ax+Bu), \\
		\hat u 	&= Wx+b.\label{eq:RC-driven2}
\end{aligned}\end{equation}

Much like the discrete time case, trajectories are generated using the above dynamics, but an arbitrary time integration scheme is necessary. The listening and training phases are conducted as usual, where the weights $W$ and biases $b$ are solved for. The prediction phase is also done by converting the nonautonomous system into an autonomous one, via the closing of the input-output loop:
\begin{equation}\begin{aligned}
		\dot x 	&= -x+\sigma(\bar A x+\bar b), \\
		\hat u 	&= Wx+b. \label{eq:RC-autonomous2}
\end{aligned}\end{equation}

Unlike in discrete time, there has yet to be a well-established and broad reaching UAP for continuous time networks. Perhaps the closest result that mimics the UAP of discrete time reservoirs traces back to the \textit{approximate realization property} of \citet{Funahashi1993}. They show that continuous time Hopfield-like networks are capable of matching the output of arbitrary dynamical systems in a window of time, but this is distinct from a claim that the the network dynamics are equivalent to the dynamics generating $u(t)$. It seemed that the progress of universal approximation in these Hopfield-like networks \eqref{eq:RC-autonomous2} had stagnated. However, with the recent introduction of NODEs \cite{chen2018neural}, interest in universal approximation of continuous time networks has steadily increased. Consider now the standard vector field of a NODE, which can be written as an autonomous reservoir with the shorthand $\bar A = A+BW$ and $\bar b = Bb$. 
\begin{equation}\begin{aligned}
		\dot x 	&= \sigma(\bar A x+\bar b), \\
		\hat u 	&= Wx+b. \label{eq:RC-autonomous3}
\end{aligned}\end{equation} 

Matrix $A$ {cannot} yet be chosen such that the ESP is \textit{a priori} guaranteed with the contraction approach following the arguments at the end of Section \ref{sec:cont-time-systems}. However, $A$ can be chosen such that the ESP occurs empirically for some input $u$ of interest. Assuming so, the reservoir trajectories are generated and the weights $W$ and bias $b$ are acquired by training. The resulting dynamics of the autonomous reservoir can be interpreted as a NODE with a linear output layer, but introduces another constraint in that the activation function are also invertible. There are multiple results \cite{Zhang2020, Teshima2020, Teshima2020a, Li2022} that have established narrower forms of UAP for NODEs, which are inherited by autonomous reservoirs \eqref{eq:RC-autonomous3} with invertible activation functions. In particular, \citet{Zhang2020} show that some augmented NODEs -- augmented by imposing that the network has number of neurons at least twice the number of inputs -- are capable of universally approximating homeomorphism of a topological space onto itself. \citet{Teshima2020} show that NODEs are universal approximators of \firstedit{a large class of} diffeomorphisms of a topological space onto itself. 

There is a significant overlap between the theory of RCs and NODEs that remains unexplored. Though the existence of the UAP in both the discrete and continuous time reservoirs are welcomed results, they are certainly not necessary nor sufficient conditions for a well-predicting reservoirs. The properties that give rise to well-predicting reservoir are an active area of research, and the gap between the necessary and sufficient conditions remains an important open problem.

\subsection{Topological Conjugacy}
Let $u$ be observations of an invertible dynamical system. If the ESP holds between $u$ and $x$, then the reservoir is synchronized (in the generalized sense) with said dynamical system. \firstedit{We will proceed assuming that $u \in U \subset \Reals^M$ are direct state observations of some system of interest, which evolves according to $f:U \to U$ on the attractor $U$. Generally speaking, direct state observations are not available, and there would be an intermediary observation function. Though the importance of observation functions cannot be understated, they are intentionally omitted to avoid additional layers of complexity. The added complexity arises from discussions around generic observables, sparse measurements, and measurement noise. We refer the reader to \citet{grigoryeva2021chaos} for a thorough treatment when observation functions are present. }

Consider the reservoirs discussed in the previous section, with the appropriate constraints such that the ESP holds and is unique. Then, there is a unique \firstedit{smooth} map $\Phi : U \to X$, induced by the dynamics of the driven reservoir, that maps the observations to the reservoir space. \firstedit{This smooth map almost surely results in an embedding, using Theorem 2.2 of \citet{sauer1991embedology}, so its inverse must also be smooth.} Corresponding to this $\Phi$, there are some autonomous dynamics $\Gamma : X \to X$ with attractor $X$ in the reservoir space. 
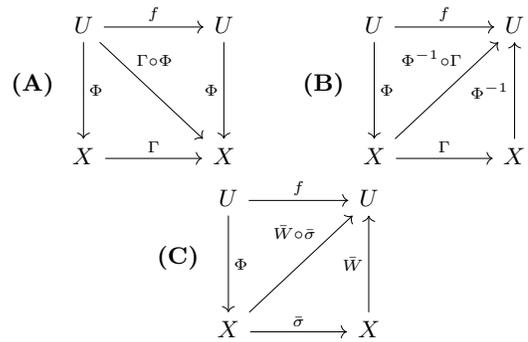
\begin{figure}
	\centering
	\textbf{(a)} \hspace{2.3cm} \textbf{(b)}\hspace{2.3cm} \textbf{(c)}\\
	\begin{tikzcd}[row sep=4em, column sep=4em]
		U \arrow[r,"f"] \arrow[d," \Phi"]\arrow[dr,"\Gamma \circ \Phi ",pos=0.3]	
		& U \arrow[d,"\Phi"']\\%
		X \arrow[r,"\Gamma"] 										
		& X 
	\end{tikzcd} \hspace{0.4cm}
	\begin{tikzcd}[row sep=4em, column sep=4em]
		U \arrow[r,"f"] \arrow[d," \Phi"]
		& U \\%
		X \arrow[r,"\Gamma"] 		\arrow[ur,"\Phi\inv \circ \Gamma ",pos=0.7]									
		& X \arrow[u,"\Phi\inv"]
	\end{tikzcd} \hspace{0.4cm}
	\begin{tikzcd}[row sep=4em, column sep=4em]
		U \arrow[r,"f"] 	\arrow[d,"\Phi"]									
		& U \\%
		X \arrow[r,"\bar \sigma"]	\arrow[ur,"\bar W \circ \bar \sigma ",pos=0.7]	
		& X \arrow[u,"\bar W"]
	\end{tikzcd} 
	
	\caption{\textbf{(a)}: The listening phase is equivalent to the composition $\Gamma \circ \Phi$. \textbf{(b)}: Corresponding to the driven reservoir, there exists a composition $\Phi\inv \circ \Gamma$. \textbf{(c)} Using the shorthand $\sigma(Ax_{t}+BWx_{t}+{Bb}) = \sigma(\bar Ax_{t}+\bar b)=\bar\sigma(x_t)$ and $Wx_t+b=\bar W(x_t)$, the autonomous reservoir has the composition $\bar W \circ \bar \sigma$ with the UAP that allows for the approximation of $\Phi\inv \circ \Gamma$ arbitrarily well. }\label{fig:conj}
\end{figure}

Since $\Phi$ \firstedit{results in an embedding, it is a homeomorphism. In the continuous time case, this would be a diffeomorphism, but we will leave it as a homeomorphism so we can discuss both discrete and continuous time reservoirs.} The dynamics $\Gamma = \Phi \circ f \circ \Phi\inv$ are topologically conjugate to $f$. The training phase of the reservoir proceeds \firstedit{under the initial assumption} that $\Phi$ is a homeomorphism, then approximates $\Phi\inv$ by an affine map with matrix $W: \Reals^{N} \to \Reals^M$ and offsets $b\in\Reals^M$. Both $W$ and $b$ are acquired using regularized least squares. As discussed around \eqref{eq:RC-autonomous}, solving for $W$ and $b$ allows for the definition of $\bar\sigma(x_t)\equiv \sigma(\bar Ax_{t}+\bar b)$, which is compatible with the UAP and an attempt to approximate the true conjugate dynamics $\Gamma$. Define the map $\bar W(x_t) = Wx_t+b$, not only does $\bar W$ approximate $\Phi\inv$, but one also gets $\bar W \circ \bar \sigma$ as an approximation to $\Phi\inv \circ \Gamma$ without additional effort. 

Despite all this, the map $\Phi$ is only induced by the dynamics when data $u$ is presented to the nonautonomous reservoir. Fig.\ref{fig:conj}\textbf{(c)} holds as an arbitrarily good approximation of Fig.\ref{fig:conj}\textbf{(b)} for the first time-step of the prediction phase. After the first step, the slightest and inevitable deviation of $\bar W \circ \bar \sigma$ from $\Phi\inv \circ \Gamma$ breaks the topological conjugacy. Such errors will compound, resulting in the eventual divergence of the initially conjugated dynamics from the original dynamics; this is distinct from the divergence due to the chaotic dynamics. In the absence of a continuous input of data, this is unavoidable save for the case of the perfect representation of $\Phi\inv \circ \Gamma$. 

Empirical evidence overwhelmingly suggests that the reservoir performs well in predicting time series of even chaotic systems. Reservoirs also maintain the `climate' of the original system and reproduces all but the most negative Lyapunov exponents \cite{Pathak2017}, which suggests that the accumulation of errors is relatively slow or possibly bounded. The robustness of the autonomous reservoir dynamics as a topological conjugate remains an open question with active and ongoing research. 

\firstedit{
\subsection{Digital Twins}

One of the most fundamental properties of RC is the necessity of contraction whilst driven by some input signal, as demonstrated throughout this paper. As long as data is continually being presented to the reservoir, the reservoir will maintain synchronization with the source system. When the data stream stops, the readout of the RC is fed back as inputs, allowing short-term predictions of the original data stream.

Interestingly, these features of the reservoir make it well suited for usage as a digital twin (DT). We will use the definition of DT put forth by the \textit{Committee on Foundational Research Gaps and Future Directions for Digital Twins} \cite{NationalAcademiesofSciences2023}, reproduced here.

\textit{``A digital twin is a set of virtual information constructs that mimics the structure, context, and behavior of natural, engineered, or social system (or system-of-systems), is dynamically updated with data from its physical twin, has a predictive capability, and informs decisions that realize value. The bidirectional interaction between the virtual and physical is central to the digital twin."}

An appropriately trained reservoir is clearly capable of mimicking certain structural aspects of the original system, most famously the invariant measures or \textit{climate} of the attractor\cite{Pathak2017}. In the listening phase, the reservoir also inherits the topology of the data due to the homeomorphism that contraction induces. To the extent that there is structure in the topology, the reservoir inherits or mimics that structure as well.

The RC framework is agnostic of the type of system that generates the data, as long as there is some low-dimensional structure in the form of an attractor. It has limited success when applied to text, speech, images, and systems governed by partial differential equations. Other than these limitations, RC has be deployed with considerable success on natural, engineered, and social systems.

The requirements of DTs to be ``dynamically updated with data from its physical twin'' and to have ``a predictive capability'' itself has striking similarities to how reservoirs operate, specifically the listening and prediction phase of RC. Though all recurrent networks can be updated with data, the presence of contraction, ESP, and GS in the RC framework is noteworthy. With the reservoir synchronized to the inputs, the topological properties of the data is guaranteed to be expressed in the reservoir, a feature that most recurrent networks do not exhibit.

There is the requirement that the virtual and physical twin have bidirectional interactions. The RC framework only allows for the development of the virtual twin from measurements of the physical twin, but leaves out the interaction on the physical using information of the virtual. RC appears to be agnostic yet entirely compatible to this segment of the DT interaction loop, so it is up to the DT's user to specify the purpose and how this can be accomplished. 

Some of these purposes, based on the strengths of RC, might be to leverage its predictive capabilities to anticipate future behavior. If future behavior is undesirable, one can intervene on the physical twin side -- say, by changing system parameters -- to avoid or correct for the undesirable behavior. Exactly how this would be accomplished is of course domain and system specific. Nonetheless, such options are open and the application space appears promising.

Lastly, there is no mention of the fidelity of the DT, defined roughly by the amount of physics in the model. In general, lower fidelity models are quicker but inherently less physically informative, while higher fidelity models are the converse. It is natural to assume that the multi-fidelity approach be desirable in DTs, since different levels of fidelity have their respective trade-offs. We are optimistic on the prospects of using RC as one of the lower fidelity components of the DT, but not as a replacements of higher fidelity and physically sound models.
}

\section{Conclusions}

This paper outlines how global nonlinear state-contraction, while driven by inputs, can be uniformly guaranteed for both discrete and continuous time reservoirs. Particularly, a discrete time reservoir with any contracting activation function $\sigma$, any connectivity matrix $A : \min_i \norm{A}_i< 1$, and any $B$ will result in the ESP and unique GS, regardless of the structure of either matrix or the inputs $u$. In continuous time, we show how the widely used Hopfield-like reservoirs with condition $\norm{A}_2 < s_N(C)$ is sufficient for contraction, before showing how lognorms can also provide sufficient conditions for the $1$-lognorm and $\infty$-lognorm cases. All these conditions induce contraction in continuous time, hence the ESP and GS hold. All of the results with matrix norms and lognorms can be further generalized with their weighted operator norms counterparts in the same manner as \citet{Buehner2006}. Towards the end of the paper, we discuss how the UAP may appear in reservoir systems and outline the overlap of continuous time reservoirs with NODEs. This suggests that the universal approximation of homeomorphisms and diffeomorphisms, proven for certain classes of NODEs, is likely inherited by continuous time reservoirs. We discuss and outline how topological conjugacy appears and remark on the surprising robustness of this conjugacy even when errors are accumulated in the prediction phase. Lastly, we lament on how the RC framework is strongly suited for DTs, particularly due to the synchronization.

\section{Acknowledgments}
The authors thank J. Koo for helpful discussions, suggestions, and encouragement during his time at the Air Force Research Laboratory. This research was supported by the Air Force Office of Scientific Research under FA9550-23RQCOR001 and FA9550-23RQCOR007.

\bibliography{Contraction-PRE}

\appendix
\clearpage\newpage
\section{Contracting Activation Function}
\label{sec:contraction-theory}

The \textit{contraction mapping property} is an incredibly useful property of a function and is repeatedly used in this paper. For readability, a function with this property will be described as \textit{contracting} or referred to as a \textit{contraction}. Let $f : \Reals^N \to \Reals^N$ be a smooth function that acts component-wise on its argument $v$, i.e. if $v = [v_1,v_2]\T$, then $f(v)=[f(v_1),f(v_2)]\T$ where the scalar- and vector-valued functions are both called $f$. This function is a contraction with respect to an arbitrarily chosen norm when the following inequality holds for $0 \le k < 1$:
\begin{equation}
		\norm{f(a)-f(b)} \le k \norm{a-b}. \label{eq:contraction1}
\end{equation}

The smallest such scalar $k$, for all $a\ne b$ in $\Reals^N$, is the \textit{Lipschitz constant} of $f$ and determines the rate at which iterated maps of $f$ converge to a unique fixed point. The value of $k$ is also equal to the magnitude of the supremum slope of $f$ in the one-dimensional setting. For the multidimensional case, the inequality \eqref{eq:contraction1} states that a contracting $f$ will always shorten the distance between any two vectors, even if a shift was performed on both vectors. Consider briefly a more general form of \eqref{eq:contraction1} by including some matrix $M\in\Reals^{N\times N}$ and offsets $c\in \Reals^N$ inside the argument. The definition of an induced matrix norm gives the following inequality:
\begin{equation}\begin{aligned}
		\norm{f(Ma+c)-f(Mb+c)} 	&\le k\norm{Ma-Mb} \\ 
		&\le k\norm{M} \!\cdot\! \norm{a-b}. \label{eq:contraction2}
\end{aligned}\end{equation}

For the above inequality to also be a contraction, consider an activation function $\sigma$ (such as the ubiquitous tanh function) that satisfies $\sigma(0)=0$, $\sigma'(0)=1$, and $\max_x \sigma'(x)=1$, where the Lipschitz constant of $\sigma$ is $k=1$. Without loss of generality, $k$ can be set to unity and contraction would occur for all vectors $a,b \in \Reals^N$ for $M$ satisfying $0 \le \norm{M} < 1$. This guarantees that $a=b$ is the unique fixed point:
\begin{equation}
		\norm{\sigma(Ma+c)-\sigma(Mb+c)} \le \norm{M} \!\cdot\! \norm{a-b}. \label{eq:contraction3}
\end{equation}

This allows for \eqref{eq:contraction3} to be invoked in \eqref{eq:discrete-contraction} to show the exponential stability of certain discrete time reservoir systems. We can also multiply the 2-norm version of \eqref{eq:contraction3} by $\norm{a-b}_2$ and apply Cauchy-Schwarz to recover the left inequality below. This inequality is used in \eqref{eq:continuous-contraction} with constraint $0 \le \norm{M} < 1$ to show the exponential stability of certain continuous time reservoir systems:
\begin{equation}
	\begin{aligned}
		\innerp{ \sigma(M&a+c)-\sigma(Mb+c), a-b} \\ &\le \norm{\sigma(Ma+c)-\sigma(Mb+c)}_2 \norm{a-b}_2 \\ &\le \norm{M}_2 \norm{a-b}_2^2. \label{eq:contraction4}
	\end{aligned}
\end{equation}

%

\section{Singular Value as Upper and Lower Bounds}\label{sec:spd-proof}

Let $C=C\T$ be a SPD matrix with an eigendecomposition $C = Q\T SQ$, where $S$ is a diagonal matrix and $Q$ is an orthonormal matrix. Being SPD, the eigenvalues of $C$ coincide with the singular values $s_n$ that populate the diagonal of $S$, ordered such that $s_1 \ge s_2 \ge \cdots \ge s_N > 0$, . 

The quantity of interest $\innerp{ Cz, z } = \norm{z\T C z}_2$ can then be written as $ \norm{(Qz)\T S (Qz)}_2 = {\sum_{n=1}^{N} s_n [Q z]_n^2}$, where $[Q z]_n$ is the $n^{\text{th}}$ component of the vector $Qz$. Since $s_n$ is ordered from largest to smallest, the inner product of interest can be written as a sum with an upper and lower bound:
\begin{equation} 
	s_N \sum_{n=1}^{N} [Q z]_n^2 \le \sum_{n=1}^{N} s_n [Q z]_n^2 \le s_1 \sum_{n=1}^{N} [Q z]_n^2.
\end{equation}

With $Q$ being orthonormal, it has no effect on the norm of $z$, i.e. $ \sum_{n=1}^{N} [Q z]_n^2 = \norm{Q z}_2^2 = \norm{z}_2^2$. The sum in both the upper and lower bounds are equal to the norm of $\norm{z}_2$. The middle term above is equal to $\innerp{ Cz, z }$ by definition, resulting in the following inequalities:
\begin{equation} 
	s_N \norm{z}_2^2 \le \innerp{ Cz, z } \le s_1 \norm{z}_2^2.
\end{equation}

\section{Properties of Logarithmic Norms}\label{sec:log-norm-prop}

The logarithmic norm (lognorm) of a square matrix $M\in\Reals^{N\times N}$ is defined as 
\begin{equation}
	\mu(A) = \lim_{h\to 0^+} \frac{\norm{I+hA}-1}{h}, \label{eq:log-norms-def}
\end{equation}

with $I$ being the appropriately sized identity matrix. The lognorm can be interpreted as 
\begin{equation}
	\begin{aligned}
		\mu(A) &= \frac{d}{dh} \norm{I+hA}\bigg|_{h\to 0^+} \\
		&= \frac{d}{dh} \norm{\exp(hA)}\bigg|_{h\to 0^+}. \label{eq:log-norms-int}
	\end{aligned}
\end{equation}

Unlike the matrix norm, the lognorm is not actually a norm. In this paper, we only make use of $\mu_1(A)$ and $\mu_\infty(A)$, which are the lognorms induced by the matrix $1$-norm and $\infty$-norm respectively. The definitions of $\mu_1(A)$ and $\mu_\infty(A)$ are the maxima of the following arguments over the columns and rows of an arbitrary square matrix $A$, with elements $a_{ij}$, respectively.

Both $\mu_1(A)$ and $\mu_\infty(A)$, shown in \eqref{eq:log-norms1} and \eqref{eq:log-norms2}, are important to guarantee the ESP and GS of continuous time reservoirs in \eqref{eq:cont-esp}. The 2-lognorm $\mu_2(A)$ has not been successfully applied in this context, but is provided here for context. Its definition is related to the spectral properties of $A+A\T$. Specifically, it is half the maximal (most positive) eigenvalue of the symmetric matrix $A+A\T$:
\begin{equation}
		\mu_2(A) = \half \lambda_{max} \left({A+A\T}\right). \\
\end{equation}

\section{Properties of Weak Pairing}\label{sec:weak-pairing-prop}
The following statements are taken from Definitions 2.17 and 2.18 of \citet{bullo2023}.
\begin{definition}A weak pairing $\weakp{ \cdot\, ,\cdot }$ is a map on $\Reals^N \times \Reals^N \to \Reals$ that satisfies:
	\begin{enumerate}[label=(\roman*)]
		\item subadditivity in the first argument $\weakp{ x_1+x_2, y } \le \weakp{ x_1, y } + \weakp{ x_2, y }$ for all $x_1,x_2,y\in\Reals^N;$
		\item continuity in the first argument $\weakp{ x, y }$ for all $x,y\in\Reals^N$;
		\item weak homogeneity $\weakp{ a x , y } = \weakp{ x , ay } = a \weakp{ x , y } $ and $\weakp{ -x , -y } = \weakp{ x , y } $ for all $x,y\in\Reals^N$ and $a\ge 0;$
		\item positive definiteness $\weakp{ x , x } > 0 $ for all $x\ne 0; $
		\item Cauchy Schwartz inequality ${\weakp{ x , y }}^2 \le \weakp{ x ,x } \weakp{ y ,y }$ for all $x,y\in\Reals^N$.
	\end{enumerate}
\end{definition}

\begin{definition}[Standing Assumptions] A weak pairing $\weakp{ \cdot\, ,\cdot }$ compatible with the norm $\norm{\cdot}$ also satisfies:
	\begin{enumerate}[label=(\roman*)]
		\item derivative formula for continuously differentiable trajectories $x(t)\in\Reals^{N}$
		$$\frac{d}{dt} \left(\half \norm{x(t)}^2 \right) =\norm{x(t)}\frac{d}{dt} \norm{x(t)} = \weakp{\dot x(t),x(t) };$$ 
		\item Lumer's equality for $A\in\Reals^{N\times N}$ and $x\in\Reals^{N}$
		$$\mu(A) = \max_{x\ne 0} \frac{ \weakp{ Ax, x }}{ \weakp{ x,x} } = \max_{\norm{x}=1} \weakp{ Ax, x };$$
		\item Deimling's inequality for $x,y\in\Reals^{N}$ 
		$${\weakp{ x , y }} \le \norm{y} \lim_{h\to 0^+} \frac{\norm{y+hx}-\norm{y}}{h}.$$
	\end{enumerate}
\end{definition}

For continuous but not differentiable trajectories, the derivative in Definition 2$(i)$ above can be replaced by the appropriate Dini derivative. 

\section{Maximum of Activation Function Jacobians}\label{sec:max-jac}

The results of this appendix are attributed to Lemma 2.15 of \citet{bullo2023}, but with additional details. To achieve a more useful form of the Jacobian $D\hat\sigma(x)$ with respect to $x$, we use the shorthand $\sigma(Ax+Bu) = \hat\sigma(x)$ and define $\sigma'$ as the derivative of the scalar activation function $\sigma$. We also define ``diag'' as the operation that places its $N$-vector argument along the diagonal of a $N\times N$ diagonal matrix, i.e. diag$(\begin{bmatrix} v_1, v_2 \end{bmatrix}\T)= \begin{bmatrix} v_1 & 0 \\ 0 & v_2 \end{bmatrix} $, and another shorthand for the resulting diagonal matrix $P = \text{diag}(\sigma'(Ax+Bu))$ with only $N$ diagonal elements $p_{jj}$:
\begin{equation}\begin{aligned}
		D\hat\sigma(x) 	&= D\sigma(Ax+Bu) \\
		&= \text{diag}(\sigma'(Ax+Bu))A \\
		&= PA \label{eq:jacobian-activation}.
\end{aligned}\end{equation}

For an element-wise activation function $\sigma$ with $\sigma'\in(0,1]$, such as the commonly used tanh function, the maximum over all $x$ can be replaced with a maximum over all diagonal elements $p_{jj}$ in the interval $(0,1]$. We first handle the 1-lognorm case:
\begin{equation}
	\begin{aligned}
		\max_{x} \mu_1(D\hat\sigma(x)) &= \max_{p_{jj}\in[0,1]} \mu(PA) \\
		&= \max_{p_{jj}\in[0,1]} \max_j \left(p_{jj}a_{jj} +\sum_{i\ne j}\abs{p_{ii}a_{ij}}\right) \\
		&= \max_j \max_{p_{jj}\in[0,1]} \left(p_{jj}a_{jj} +\sum_{i\ne j}\abs{p_{ii}a_{ij}}\right) \\
		&= \max_j 
		\begin{cases}
			a_{jj}+\sum_{i\ne j}\abs{a_{ij}}, & \text{for } a_{jj}\ge 0\\
			0+\sum_{i\ne j}\abs{a_{ij}}, & \text{for } a_{jj} < 0
		\end{cases} \\
		&= \max 
		\begin{cases}
			\max_j a_{jj}+\sum_{i\ne j}\abs{a_{ij}} \\
			\max_j \sum_{i\ne j}\abs{a_{ij}} 
		\end{cases} \\
		&= \max\{ \mu_1(A) , \mu_1(A-I\circ A) \}.
	\end{aligned}
\end{equation}

The matrix $(A-I\circ A)$ makes use of the entry-wise or Hadamard product, where the elements $(I\circ A)_{ij} = (I)_{ij}(A)_{ij}$. More simply, $(A-I\circ A)$ is just $A$ with its diagonal elements set to zero. Next, we handle the $\infty$-norm case. This case differs from the 1-norm case because the elements of the diagonal matrix $P$ act on the rows of $A$, which allow for $p_{ii}$ to be factored out of the row sum. We define yet another shorthand $q_i = a_{ii} +\sum_{j\ne i}\abs{a_{ij}}$:
\begin{equation}
	\begin{aligned}
		\max_{x} \mu_\infty(D\hat\sigma(x)) &= \max_{p_{ii}\in[0,1]} \mu_\infty(PA) \\
		&= \max_{p_{ii}\in[0,1]} \max_i \left(p_{ii}a_{ii} +\sum_{j\ne i}\abs{p_{ii}a_{ij}}\right) \\
		&= \max_i \max_{p_{ii}\in[0,1]} p_{ii} q_i \\
		&= \max_i 
		\begin{cases}
			q_i, & \text{for } q_i\ge 0\\
			0, & \text{for } q_i < 0
		\end{cases} \\
		&= \max\{ \mu_\infty(A) , 0 \}.
	\end{aligned}
\end{equation}

Lastly, we provide some bounds using the equivalent matrix norms. The matrix 1-norm is defined as the maximum of the absolute \textbf{column} sum and the matrix $\infty$-norm is defined as the maximum of the absolute \textbf{row} sum:
\begin{equation}
	\norm{A}_1 = \max_j \left(\sum_{i}\abs{a_{ij}}\right), \\
\end{equation}
\begin{equation}
	\norm{A}_\infty = \max_i \left(\sum_{j}\abs{a_{ij}}\right).
\end{equation}

The lognorms, defined in \eqref{eq:log-norms1} and \eqref{eq:log-norms2}, are bounded by the equivalent matrix norms because the lognorms allow the diagonal elements to be negative. We have $\mu_1(A) \le \norm{A}_1$ and $\mu_\infty(A) \le \norm{A}_\infty$. Additionally, $\mu_1(A-I\circ A) \le \norm{A}_1$ because $\mu_1(A-I\circ A)$ omits the diagonal elements in the sum entirely.

\end{document}